\documentclass[jmp,aps,amsmath,amssymb,floatfix]{revtex4-1}

\usepackage{graphicx}
\usepackage{dcolumn}
\usepackage{bm} 
\usepackage{amssymb}
\usepackage{epsfig}
\usepackage{color}

\newcommand{\ba}{\begin{eqnarray}}
\newcommand{\ea}{\end{eqnarray}}
\newcommand{\be}{\begin{equation}}
\newcommand{\ee}{\end{equation}}
\newcommand{\bdisplay}{\begin{displaymath}}
\newcommand{\edisplay}{\end{displaymath}}

\newcommand{\eq}[1]{Eq.\,(\ref{#1})}

\begin{document}

\title{Asymptotic Bessel-function expansions for Legendre and Jacobi functions}

\author{Loyal Durand}
\email{ldurand@hep.wisc.edu}
\altaffiliation{Mailing address: 415 Pearl Ct., Aspen, CO 81611}
\affiliation{Department of Physics, University of Wisconsin, Madison, WI 53706 USA}

\begin{abstract}

We present new asymptotic series for the Legendre and Jacobi functions of the first and second kinds in terms of Bessel functions with appropriate arguments. The results are useful in the context of scattering problems, improve on known limiting results, and allow the calculation of corrections to the leading Bessel-function approximations for these functions. Our derivations of these series are based on Barnes-type representations of the Legendre, Jacobi, and Bessel functions; our method appears to be new. We use the results, finally, to obtain asymptotic Bessel function expansions for the rotation functions needed to describe the scattering of particles with spin.
 
\end{abstract}
 
\maketitle


\section{Introduction \label{sec:introduction}}

Legendre and Jacobi functions of the first and second kind often appear in scattering theory in contexts in which they can be approximated by Bessel functions of appropriate arguments. For example, the scattering amplitude for the elastic scattering two spinless equal-mass particles can be expanded in the partial-wave series 
\be
\label{f_elas}
f(s,t) = \sum_j (2j+1)f_j(s)P_j(\cos{\theta}).
\ee
Here $j$ is the angular momentum, $\theta$ is the scattering angle, $s=E^2=4(p^2+m^2)$ is the square of the energy, and $t=2p^2(\cos{\theta}-1)=-q^2$ is the invariant momentum transfer,  with $p$ the three-momentum of the particles in the center-of-mass system.  The partial-wave scattering amplitudes are $f_j(s,t)=(e^{2i\delta_j}-1)/2ip$ where $\delta_j(E)$ is the complex phase shift for scattering in the $j^{\rm th}$ partial wave. We will write this as $2\delta_j(E)=\chi(j,E)=\chi^R(j,E)+i\chi^I(j,E)$.  The elastic scattering cross section is  $d\sigma_{\rm elas}/d\Omega=|f(s,t)|^2$ and the total cross section is  $\sigma_{\rm tot}=(4\pi/p)\,{\rm Im}\,f(s,0)$ by the optical theorem.

 At high energies, the scattering is generally strongly inelastic with $\chi^I(j,E)$ large. The scattering involves many angular momenta---{\em e.g.},  $j\sim 80$ at the peak of the $j$ distribution for the total cross section in  $pp$ scattering at $E= 50$ GeV, with significant contributions up to $j\sim 250$---with  $\chi(j,E)$  varying smoothly with $j$.  We will  therefore treat $j$ as a continuous variable. 
 
 The scattering is also strongly peaked in the forward direction, $\theta\ll 1$, which allows us to use the well-known small-angle Bessel-function approximation for the Legendre functions  \cite{watson},\, Sec.\,5.71, and \cite{HTF},\,(HTF2,\,7.8(2)), 
 \be
 \label{Pjsmall}
 P_j(\cos{\theta}) \approx J_0\left(\sqrt{2j(j+1)(1-\cos{\theta}})\right) = J_0(b\sqrt{-t}),
 \ee
where  $b=\sqrt{j(j+1)}/p$  is the radial turning point or impact parameter for the free-particle Schr{\"{o}}dinger equation. Then with $\sum_j (2j+1)\rightarrow 2p^2\int bdb$, we can then convert the sum over $j$ to an integral, and  obtain the impact-parameter or eikonal  representation of the scattering amplitude \cite{blockrev},
\be
\label{f_defined}
f(s,t) \approx ip\int_0^\infty db\,b\left(1-e^{i\chi(b,s)}\right)J_0\left(b\sqrt{-t}\right).
\ee

In the case of particles with spin, the partial-wave scattering amplitudes are helicity dependent \cite{jacob_wick}, and the Legendre functions are replaced by the rotation coefficients $d^j_{m',m}(\cos{\theta})$ \cite{Edmonds,Rose}, where $m'$ and $m$ are the total initial and final helicity projections. The rotation coefficients are expressible in terms of the Jacobi polynomials $P_{j-m'}^{(m'-m,m'+m)}(\cos{\theta})$. These again have a small-angle approximation in terms of Bessel functions \cite{szego}, Sec.\ 8.1,, allowing us to obtain the spin-dependent analog of \eq{f_defined} as, {\em e.g.}, in \cite{LD_chiu1965}.

It is important for such applications to determine the accuracy of the small-angle approximations---\eq{Pjsmall} and its analogs for the other Legendre and Jacobi functions---and to obtain systematic estimates for the errors. MacDonald (HTF1,\,3.5(10))  has given an asymptotic expansion for the Legendre functions $P_j^\mu(\cos{\theta})$ (HTF1,\,3.5(10)) in terms of Bessel functions and powers of their arguments which is useful in this context. (The more precise uniform expansion given by Szeg{\" o} (HTF2,\,7.8(15)) involves extra trigonometric functions, and is less useful.) Much less is apparently known about Bessel function expansions for the general Jacobi functions $P_j^{(\alpha,\beta)}(x)$ or the corresponding Legendre and Jacobi functions of the second kind, $Q_j^\mu(x)$ and $Q_j^{(\alpha,\beta)}(x)$, all of which are needed in scattering problems as discussed, for example, in \cite{LD_chiu1965} and \cite{AndrewsGunson}.

Our objective here is to obtain  asymptotic Bessel function expansions of all these functions which are useful in the context of scattering theory. Our method, based on Barnes-type representations \cite{WhittakerWatson},\,14.5, of the Legendre, Jacobi, and Bessel functions, is apparently new. 

We begin in Sec.\ \ref{subsec:Pj} with the Legendre functions $P_j^\mu(x)$, develop our method in detail, and obtain a Bessel-function expansion similar to MacDonald's, but somewhat better because of a better choice of the expansion parameter. We then extend the results to the functions of the second kind, $Q_j^\mu(x)$, in Sec.\ \ref{subsec:Qj}, and to the functions ${\mathsf P}_j^\mu(x)$ and ${\mathsf Q}_j^\mu(x)$ ``on the cut'', $-1<x<1$, in Sec.\ \ref{Legendre_cut}. 

We further extend our results to the general Jacobi functions $P_j^{(\alpha,\beta)}(x)$ and $Q_j^{(\alpha,\beta)}(x)$ in Secs.\  \ref{subsec:Jacobi_P},  \ref{subsec:Jacobi_Q}, and \ref{subsecJacobi_on_cut}, and, finally, to the rotation functions $d^j_{m',m}(x)$ and $e^j_{m',m}(x)$ in Sec.\ \ref{rotation_coeff}.


\section{Bessel function approximations for Legendre functions \label{sec:Legendre_funcs}} 


\subsection{The functions $P_j^\mu(x)$ \label{subsec:Pj}}

The Legendre functions of the first kind,
\be
\label{Legendre2F1}
P_j^{-\mu}(x)=\frac{1}{\Gamma(1+\mu)}\left(\frac{x-1}{x+1}\right)^{\mu/2} \ _2F_1\left(-j,j+1;1+\mu;\frac{1-x}{2}\right),
\ee
have the Barnes-type integral representation (HFT1, 2.1.3(15) and 3.2(14))  \cite{HTF,WhittakerWatson}
\be
\label{PjBarnes}
P_j^{-\mu}(x) =\left(\frac{x-1}{x+1}\right)^{\mu/2} \frac{1}{2\pi i}\int_{-i\infty}^{i\infty}ds\frac{\Gamma(-j+s)\Gamma(j+1+s)}{\Gamma(-j)\Gamma(j+1)}\frac{\Gamma(-s)}{\Gamma(s+\mu+1)}\left(-\frac{1-x}{2}\right)^s.
\ee
We will assume initially that $j$ is non-integer and ${\rm Re}\mu>0$. The integration contour in $s$ runs from $-i\infty$ to $+i\infty$, and is distorted to pass the poles in the functions $\Gamma(-j+s)$ and $ \Gamma(j+1+s)$ on the right, and those of $\Gamma(-s)$ on the left. We will take the limit of physical integer values of $j$ and vanishing $\mu$ later.

The Bessel functions $J_\nu(z)$ have a similar Barnes-type integral representation for $z$ real and positive and ${\rm Re}\,\nu>0$ \cite{watson}, 6.5(7),
\be
\label{Bessel_int}
J_\nu(z) = \frac{1}{2\pi i}\int_{-i\infty}^{i\infty}ds \frac{\Gamma(-s)}{\Gamma(\nu+s+1)}\left(\frac{z}{2}\right)^{\nu+2s}.
\ee
We will exploit this similarity in what follows.

We begin by expanding the ratio of gamma functions $\Gamma(-j+s)\Gamma(j+1+s)\big/\Gamma(-j)\Gamma(j+1)$
 in \eq{PjBarnes} in an asymptotic series using Stirling's approximation for $\ln\Gamma(x)$, assuming that $|j|\gg |s|$.  This condition is certainly not true on the entire contour of integration in \eq{PjBarnes}, so the series we obtain by integrating the initial expansion term-by-term will be at best asymptotic. 

The initial expansion gives an overall factor $(-j(j+1))^s$ multiplying a series in inverse powers of $j(j+1)$, with mixed powers of $s$ in the numerators. We rearrange the latter into sums of terms of the form $s(s-1)\cdots(s-k)$. Collecting terms up to $k=5$ with inverse powers of $j(j+1)$ up to the fourth, we get
\ba
\frac{\Gamma(-j+s)\Gamma(j+1+s)}{\Gamma(-j)\Gamma(j+1)} &\approx& \left(-j(j+1)\right)^s\left[1-\frac{s(s-1)}{j(j+1)}-\left(\frac{1}{3j(j+1)}-\frac{2}{j^2(j+1)^2}\right)s(s-1)(s-2) \right. \nonumber \\
&& +\left(\frac{5}{2j^2(j+1)^2}-\frac{6}{j^3(j+1)^3}\right)s(s-1)(s-2)(s-3) \nonumber \\
&& \left. +\left(\frac{11}{15j^2(j+1)^2}-\frac{66}{5j^3(j+1)^3}+\frac{24}{j^4(j+1)^4}\right)s(s-1)\cdots (s-4) \right. \nonumber \\
\label{gamma_expansion}
&& \left.+ \left(\frac{1}{18j^2(j+1)^2}-\frac{49}{6j^3(j+1)^3}+\frac{76}{j^4(j+1)^4}+\cdots\right)s(s-1)\cdots(s-5)+\cdots\right].
\ea 

We will use this expansion in \eq{PjBarnes}, defining a new variable $z$ with $z=\sqrt{2j(j+1)(1-x)}$  and combining the factors $s(s-1)\cdots(s-k)$ with $\Gamma(-s)$ to get $(-1)^{k+1}\Gamma(-s+k+1)$. We can then shift the contour of the $s$ integration to the right, to run from $k+1-i\infty$ to $k+1+i\infty$ passing just to the left of the pole at $s=k+1$. Replacing $s$ by $s'=s-k-1$, we find that 
\ba
&& \frac{1}{2\pi i}\int_{-i\infty}^{i\infty} ds\,s(s-1)\cdots(s-k)\frac{\Gamma(-s)}{\Gamma(s+\mu+1)}\left(j(j+1)\frac{1-x}{2}\right)^s \nonumber \\
&& \qquad = \frac{1}{2\pi i}(-1)^{k+1} \int_{-i\infty}^{i\infty}ds'\,\frac{\Gamma(-s')}{\Gamma(s'+\mu+k+2)}\left(\frac{z}{2}\right)^{2s'+2k+2} \nonumber \\
&& \qquad = (-1)^{k+1}(z/2)^{k+1-\mu}J_{k+1+\mu}(z).
\label{J_(k+1)}
\ea

This result is general. We can go to integer values of $j$ without difficulty, and can continue the result to general, possibly complex, values of $\mu$ and $x$ or $z$ by distorting the integration contour in $s'$ to run from $-\infty$ to $+\infty$ parallel to the real axis staying just to the left of $s'=0$ as before. 

Using this result with \eq{gamma_expansion}, we obtain an asymptotic series for the Legendre function $P_j^{-\mu}(x)$ in terms of Bessel functions $J_{k+1+\mu}$,
\ba
P_j^{-\mu}(x) &\approx& \left(\frac{x-1}{x+1}\right)^{\mu/2}\left\{\left(\frac{z}{2}\right)^{-\mu}J_\mu(z)-\frac{1}{j(j+1)}\left(\frac{z}{2}\right)^{2-\mu}J_{2+\mu}(z)  \right. \nonumber \\
&& + \left(\frac{1}{3j(j+1)}-\frac{2}{j^2(j+1)^2}\right)\left(\frac{z}{2}\right)^{3-\mu} J_{3+\mu}(z)  \nonumber \\
&& +\left(\frac{5}{2j^2(j+1)^2}-\frac{6}{j^3(j+1)^3}\right)\left(\frac{z}{2}\right)^{4-\mu} J_{4+\mu}(z) \nonumber \\
&& - \left(\frac{11}{15j^2(j+1)^2}-\frac{66}{5j^3(j+1)^3}+\frac{24}{j^4(j+1)^4}\right)\left(\frac{z}{2}\right)^{5-\mu} J_{5+\mu}(z) \nonumber \\
&& +\left(\frac{1}{18j^2(j+1)^2}-\frac{49}{6j^3(j+1)^3}+\frac{76}{j^4(j+1)^4}+\cdots\right)\left(\frac{z}{2}\right)^{6-\mu}J_{6+\mu}(z) \nonumber \\
\label{Jseries1}
&& \left. +O\left(\frac{1}{j^3(j+1)^3}\left(\frac{z}{2}\right)^7 J_7(z)\right)\right\}.
\ea

Since $j$ is typically large in applications, we will rearrange the series in a form which makes it possible to estimate the uncertainties systematically,
\ba
P_j^{-\mu}(x) &\approx&  \left(\frac{z'}{2}\right)^{-\mu}\left\{ J_\mu(z)-\frac{1}{j(j+1)}\left[\left(\frac{z}{2}\right)^{2}J_{2+\mu}(z)-\frac{1}{3}\left(\frac{z}{2}\right)^{3}J_{3+\mu}(z)\right] \right. \nonumber \\
&& -\frac{1}{j^2(j+1)^2}\left[2\left(\frac{z}{2}\right)^{3}J_{ 3+\mu}(z) -\frac{5}{2}\left(\frac{z}{2}\right)^{4}J_{4+\mu}(z) \right. \nonumber \\
\label{Pjseries1}
&& \left.\left. +\frac{11}{15}\left(\frac{z}{2}\right)^{5}J_{5+\mu}(z)-\frac{1}{18}\left(\frac{z}{2}\right)^{6}J_{6+\mu}(z) \right]+\cdots \right\}
\ea
where, again, $ z=\sqrt{2j(j+1)(1-x)}$, and $z'=\sqrt{2j(j+1)(1+x)}$. The following terms are of order $1/j^3(j+1)^3$ and higher. The leading factor can be expanded for $z$ fixed and $j$ large as
\be
\label{z'_expanded}
\left(\frac{z'}{2}\right)^{-\mu} = 1 +\frac{\mu}{j(j+1)}\left(\frac{z}{2}\right)^2 + \frac{1}{2}\frac{\mu(\mu+1)}{j^2(j+1)^2}\left(\frac{z}{2}\right)^4 + \cdots,
\ee
so differs from 1 only to order $1/j(j+1)$.

We note also that, if we retain terms in \eq{Pjseries1}, regarded as an expansion in Bessel functions, through $(z/2)^{n-\mu}J_\mu(z)$, the result reproduces the hypergeometric expansion of $P_j^{-\mu}(x)$ in powers of $(1-x)$ through order $n$. As a result, the series also holds for $j$ small provided $(1-x )$ is small enough.
  
For the common case $\mu=0$, with $P_j(x)\equiv P_j^0(x)$ the series reduces to
\ba
P_j(x) &\approx&  J_0(z)-\frac{1}{j(j+1)}\left[\left(\frac{z}{2}\right)^2J_2(z)-\frac{1}{3}\left(\frac{z}{2}\right)^3J_3(z)\right] \nonumber \\
\label{Pj_series}
&& -\frac{1}{j^2(j+1)^2}\left[2\left(\frac{z}{2}\right)^3J_3(z) -\frac{5}{2}\left(\frac{z}{2}\right)^4J_4(z) 
+\frac{11}{15}\left(\frac{z}{2}\right)^5J_5(z)-\frac{1}{18}\left(\frac{z}{2}\right)^6J_6(z) \right]+\cdots . 
\ea
This series is similar in form to MacDonald's series for $P_j(x)$ in inverse powers of $\left(j+\frac{1}{2}\right)$ (HTF1, 3.5(10)),
\be
\label{MacDonald}
P_j(x) \approx J_0(z'') + \frac{1}{(j+\frac{1}{2})^2}\left[\frac{1}{4}\left(\frac{z''}{2}\right)J_1(z'')-\left(\frac{z''}{2}\right)^2J_2(z'')+\frac{1}{3}\left(\frac{z''}{2}\right)^3J_3(z'')\right]+\cdots,
\ee
where $z''=(j+\frac{1}{2})\sqrt{2(1-x)}$.  However, our use of $j(j+1)$ instead of $j+\frac{1}{2}$ as the expansion parameter eliminates the leading correction term proportional to $J_1(z'')$ in MacDonald's expansion. The $J_1(z'')$ term and similar terms later in MacDonalds series result from the expansion of the factor $j(j+1)=(j+\frac{1}{2})^2-\frac{1}{4}$ in the argument $z$ in \eq{Pj_series} in terms of $(j+\frac{1}{2})$. 

The difference between the two approximations is small for $j\gg 1$ when the expansion parameter is taken as $1/\sqrt{j(j+1)}\approx 1/( j+\frac{1}{2})$ with $z$ taken as fixed. If, however,  one considers the relevant expansion parameter as $\sqrt{(1-x)/2}=\sin(\theta/2)$, $x=\cos{\theta}$---that is, treats the series as a small-angle expansion with the Bessel functions treated as of $O(1)$---the leading correction terms are $O(\sin^2(\theta/2))$ for our series, but only $O(\sin(\theta/2))$ for MacDonald's, a significant difference for applications and when when estimating errors.

In the context of scattering theory, we can regard the small-angle expansionas an expansion in powers of the ratio of the momentum transfer $q$ to the initial momentum $p$. Expressed in terms of $q$, $p$, and the impact parameter $b$, \eq{Pjseries1} becomes
\be
\label{qp_series}
P_j^{-\mu}(x) \approx  \left(\frac{z'}{2}\right)^{-\mu}\left\{ J_\mu(qb)-\frac{q^2}{p^2}\left[J_{2+\mu}(qb)-\frac{1}{3}\left(\frac{qb}{2}\right)J_{3+\mu}(qb)\right] + O(q^4/p^4) \right\},
\ee
a form which makes it clear that ``small angle'' means $q^2/p^2\ll 1$.

The asymptotic character of the expansions in Eqs.\ (\ref{Jseries1}) and (\ref{Pj_series}) noted after \eq{Bessel_int}  is clear. There are two relevant quantities in the Bessel function approximation for $P_j(x)$, namely $j(j+1)$ and $(1-x)$.  Successive terms in the series in \eq{Pj_series} initially decrease rapidly with increasing values of $j$  provided that $z=\sqrt{2(1-x)j(j+1)}$ is kept fixed and $O(1)$, but grow with powers of $j$ for increasing $j$ and fixed $x$. Alternatively, with $j$ large and $z$ fixed, we obtain a series in powers of $(1-x)$, or, in the scattering context, powers of $q^2/2p^2$, which initially converges rapidly for $(1-x)$ or $q^2/2p^2$ small. As we noted earlier, the expression in \eq{Jseries1} reproduces the hypergeometric expansion of  $P_j^\mu(x)$ in powers of $(1-x)$ exactly through terms of order $(1-x)^n$ when the Bessel functions through $J_{n+\mu}(z)$ are retained, but with errors for larger powers of $(1-x)$ that increase with $j$. 

The expansions above are clearly most useful for $j$ large and $(1-x)$ small simultaneously. However, in contrast to MacDonald's expansion, they are also well-behaved for $j$ small, giving the proper limit $P_j^{-\mu}(x)\rightarrow \left((x-1)/(x+1)\right)^{\mu/2}/\Gamma(\mu+1)$ for $j\rightarrow 0$.

Our derivation of \eq{Pjseries1} assumed that $|(1-x)/2|< 1 $ for $x$ real or complex.
We  can continue  that expression to the region with $|x-1|/2>1$, in particular to $z$ real, $z=\cosh{\theta}>1$, in terms of the hyperbolic Bessel functions $I_j^\mu(x)$ of the first kind,
\be
\label{J_mu-->I_mu}
\left(\frac{z}{2}\right)^{n-\mu}J_{n+\mu}(z) \rightarrow (-1)^n\left(\frac{Z}{2}\right)^{n-\mu} I_{n+\mu}(Z), \quad Z=\sqrt{2j(j+1)(x-1)}.
\ee
Rewriting the prefactor $\left[(x-1)/(x+1)\right]^{\mu/2}$ in \eq{Pjseries1} as $(Z/2)^{\mu/2}(Z'/2)^{-\mu/2}$ with $Z'=\sqrt{2j(j+1)(x+1)}$, we obtain
\ba
P_j^{-\mu}(x) &\approx&  \left(\frac{Z'}{2}\right)^{-\mu/2} \left\{I_\mu(Z) -\frac{1}{j(j+1)}\left[\left(\frac{Z}{2}\right)^{2}I_{2+\mu}(Z)+\frac{1}{3}\left(\frac{Z}{2}\right)^{3}I_{3+\mu}(Z)\right] \right. \nonumber \\
\label{I_series}
&& \left. + \frac{1}{j^2(j+1)^2}\left[2\left(\frac{Z}{2}\right)^{3}I_{ 3+\mu}(Z)+\cdots\right]+O\left(\frac{1}{j^3(j+1)^3}\right)\right\}, \quad x>1.
\ea
The series is again asymptotic.


\subsection{The functions $Q_j^\mu(z)$ \label{subsec:Qj}}

The Legendre functions of the second kind, $Q_j^\mu(x)$, are given by (HTF1, 3.2(32,\,37))
\ba
\label{Qjbasic}
Q_j^\mu(x) &=& \frac{1}{2}e^{i\pi\mu}\frac{\Gamma(j+1)\Gamma(j+\mu+1)}{\Gamma(2j+2)}\left(\frac{x-1}{2}\right)^{-j-\frac{1}{2}\mu-1}\left(\frac{x+1}{2}\right)^{\frac{1}{2}\mu}\  _2F_1\left(j+\mu+1,j+1;2j+2;\frac{2}{1-x}\right) \\
&=& e^{i\pi \mu} \frac{\pi}{2\sin{\pi\mu}}\left[\frac{1}{\Gamma(1-\mu)}\left(\frac{x-1}{x+1}\right)^{-\mu/2}\ _2F_1\left(-j,j+1;1-\mu; \frac{1-x}{2}\right) \right. \nonumber \\
\label{Qj_2F1}
&& \left. -\frac{1}{\Gamma(1+\mu)}\frac{\Gamma(j+\mu+1)}{\Gamma(j-\mu+1)}\left(\frac{x-1}{x+1}\right)^{\mu/2}\ _2F_1\left(-j,j+1;1+\mu;\frac{1-x}{2}\right) \right] \\
\label{QjPjform}
&& = e^{i\pi \mu} \frac{\pi}{2\sin{\pi\mu}}\left[ P_j^\mu(x) - \frac{\Gamma(j+\mu+1)}{\Gamma(j-\mu+1)} P_j^{-\mu}(x)\right].
\ea
The first relation holds for for $|x-1|/2>1$, the second for for $|x-1|/2<1$. We will initially consider the second case.

Upon substituting  the expansion in \eq{I_series} in \eq{QjPjform}, we obtain an asymptotic series for $Q_j^\mu(x)$,
\ba
Q_j^\mu(x) &=& e^{i\pi \mu} \frac{\pi}{2\sin{\pi\mu}}\left\{ \left[\left(\frac{Z'}{2}\right)^{\mu}I_{-\mu}(Z) - \frac{\Gamma(j+\mu+1)}{\Gamma(j-\mu+1)}\left(\frac{Z'}{2}\right)^{-\mu} I_{\mu}(Z) \right] \right. \nonumber \\
&& \left. -\frac{1}{j(j+1)}\left[\left(\frac{Z'}{2}\right)^\mu\left(\frac{Z}{2}\right)^2 I_{2-\mu}(Z)-\frac{\Gamma(j+\mu+1)}{\Gamma(j-\mu+1)} \left(\frac{Z'}{2}\right)^{-\mu} \left(\frac{Z}{2}\right)^2I_{2+\mu} \right. \right.\nonumber \\
\label{QJ_general}
&& \left.\left.  +\frac{1}{3}\left(\frac{Z'}{2}\right)^\mu\left(\frac{Z}{2}\right)^3 I_{3-\mu}(Z)
-\frac{1}{3}\frac{\Gamma(j+\mu+1)}{\Gamma(j-\mu+1)} \left(\frac{Z'}{2}\right)^{-\mu} \left(\frac{Z}{2}\right)^3 I_{3+\mu}(Z)  
\right]+\cdots\right\}
\ea
where  $Z=\sqrt{2j(j+1)(x-1)}$ as above,  $Z'=\sqrt{2j(j+1)(x+1)}$, and the next erm in the series is $O(1/(j(j+1))^2$.

This is the general result. We note, however, that the Bessel functions $I_{n-\mu}(Z)$ with $\mu>n$ are singular for $Z\rightarrow 0$, and can usefully be written for some purposes in terms of the hyperbolic Bessel functions $K_{\mu-n}(Z)$ and $I_{\mu-n}(Z)$ using the relation  \cite{watson}\,3.7(4), 
\be 
\label{K_nu_defined}
K_\nu(x) = \frac{\pi}{2}\,\frac{I_{-\nu}(x)-I_\nu(x)}{\sin{\nu\pi}}.
\ee
In the limit $\nu= 0$, this becomes
\be
\label{K_nu_derivative}
K_0(x) = \frac{1}{2}\lim_{\nu\rightarrow 0}\frac{d}{d\nu}\left[I_{-\nu}(x)-I_\nu(x)\right].
\ee
Thus, for the leading term, we have 
\ba
&&\frac{\pi}{2}\frac{1}{\sin{\pi\mu}}e^{i\pi\mu}\left(\frac{Z'}{2}\right)^{\mu}\left[I_{-\mu}(Z) - \frac{\Gamma(j+\mu+1)}{\Gamma(j-\mu+1)}\left(\frac{Z'}{2}\right)^{-2\mu} I_{\mu}(Z) \right] \nonumber \\
&& \qquad  = e^{i\pi\mu}\left(\frac{Z'}{2}\right)^\mu \left[K_\mu(Z)+\frac{\pi}{2}\frac{1}{\sin{\pi\mu}} I_{\mu}(Z)\left(1-(j(j+1))^{-\mu}\frac{\Gamma(j+\mu+1)}{\Gamma(j-\mu+1)} \left(\frac{x+1}{2}\right)^{-\mu}\right)\right],
\ea
where we have  assumed that ${\rm Re}\mu\geq 0$ in writing $I_{-\mu}(Z)$ in terms of $K_\mu(Z)$ and $I_\mu(Z)$. 

We can expand the coefficient of $I_\mu(Z)$ in the second line for $j\gg 1$ and fixed $Z$ using Stirling's approximation for the $\Gamma$ functions and expressing $(x+1)/2$ as $1+(Z/2)^2/j(j+1)$, and find that the leading term in the asymptotic expression for $Q_j^\mu(x)$ is
\be
\label{n=0_term}
 e^{i\pi\mu}\left(\frac{Z'}{2}\right)^\mu \left[K_\mu(Z)+\frac{\pi}{2}\frac{\mu}{\sin{\pi\mu}}\left(\frac{1}{j(j+1)}\left(\frac{Z}{2}\right)^2-\frac{1}{3(j+1)^2}\right)I_\mu(Z)+O\left(\frac{1}{(j+1)^3}\right)\right].
 \ee
 The remaining terms from  \eq{I_series} with $n\geq 2$ are explicitly of order $1/j(j+1)$ or smaller. 
 
 We find, therefore, that
 \be
 \label{Qj^mu_limit}
 Q_j^\mu(x) \sim e^{i\pi\mu}\left(\frac{Z'}{2}\right)^\mu K_\mu(Z)+O\left(\frac{1}{j(j+1)}\right)
 \ee
 for $j$ large and $Z=\sqrt{2j(j+1)(x-1)}$ fixed. We can treat the higher-order terms similarly, but the results become increasingly complicated.

In the important limiting case $\mu\rightarrow 0$, $Q_j^\mu(x)\rightarrow Q_j(x)$ needed in common applications, the orders $n\pm\mu$ on the Bessel functions in \eq{QJ_general} can be reduced to $\pm\mu$ using the relations
\be
\label{I_(n+mu)}
I_{n\pm\mu}(Z)=Z^{n\pm\mu}\left(\frac{1}{Z}\frac{d}{dZ}\right)^n Z^{\mp\mu}I_{\pm\mu}(Z),\quad K_{n\pm\mu}(Z)=Z^{n\pm\mu}\left(-\frac{1}{Z}\frac{d}{dZ}\right)^n Z^{\mp\mu}K_{\pm\mu}(Z).
\ee

We can then use the relation in \eq{K_nu_defined} to reduce this expression to one in terms of $K_\mu(Z)$ and $I_\mu(z)$.   In the limit $\mu=0$, we obtain just the derivative of the bracketed term in \eq{QJ_general} with respect to $\mu$,
\ba
&& \frac{1}{2}\frac{d}{d\mu}\left[\left(\frac{Z'}{2}\right)^{\mu}\left(\frac{Z}{2}\right)^n\cdot Z^{n-\mu}\left(\frac{1}{Z}\frac{d}{dZ}\right)^n Z^\mu I_{-\mu}(Z) \right. \nonumber \\
&& \quad \left. -  \frac{\Gamma(j+\mu+1)}{\Gamma(j-\mu+1)}\left(\frac{Z'}{2}\right)^{-\mu}\left(\frac{Z}{2}\right)^n\cdot Z^{n+\mu}\left(\frac{1}{Z}\frac{d}{dZ}\right)^n Z^{-\mu} I_\mu(Z) \right]_{\mu=0} \nonumber \\
\label{reduced_type_term}
&& = \left(-\frac{Z}{2}\right)^nK_n(Z)+\left(\frac{Z}{2}\right)^nI_n(Z)\left[\ln{\frac{Z'}{2}}-\psi(j+1)\right]-\frac{1}{2}\sum_{k=1}^n\frac{(-1)^k}{k}\frac{\Gamma(n+1)}{\Gamma(n-k+1)}\left(\frac{Z}{2}\right)^{n-k}I_{n-k}(Z),
\ea
where $\psi(j+1)$ is the digamma function.

We obtain the final result for $Q_j(x)\equiv Q_j^0(x)$ by combining these expressions with the coefficients in \eq{I_series}. The results are rather complicated but present no difficuties. To the explicit order $1/j(j+1)$,
\ba
\label{Qj_K0}
Q_j(x) &\approx& K_0(Z)+I_0(Z)\left(\ln{\frac{Z'}{2}}-\psi(j+1)\right)-\frac{1}{j(j+1}\left[\left(\frac{Z}{2}\right)^2K_2(Z)-\frac{1}{3}\left(\frac{Z}{2}\right)^3K_3(Z)\right.\nonumber \\
&& +\left(\left(\frac{Z}{2}\right)^2I_2(Z)+\frac{1}{3}\left(\frac{Z}{2}\right)^3I_3(Z)\right)\left(\ln{\frac{Z'}{2}}-\psi(j+1)\right) \nonumber \\
\label{Qj_K0}
&& \left. -\frac{1}{6}I_0(Z)+\frac{1}{2}\left(\frac{Z}{2}\right)I_1(Z)+\frac{1}{2}\left(\frac{Z}{2}\right)^2I_2(Z) \right] +O\left(\frac{1}{j^2(j+1)^2}\right).
\ea

We note that the terms in this equation proportional to $\ln{(Z'/2)}$ reflect for the expected logarithmic singularity of $Q_j(x)$ at $x=-1$, $Z'=0$, which is not included in the functions $K_n(Z)$; this should not be relevant in the region of interest,  $j$ large and $x\approx 1$. In fact, the combination in which it appears, $\ln{(Z'/2)}-\psi(j+1)$, {\em e.g.}, as the coefficient of $I_0(Z)$ in the first line of \eq{Qj_K0}, is of order $1/j(j+1)$,
\ba
 \ln(Z'/2)-\psi(j+1) &=& \frac{1}{2}\ln\left(1+\frac{1}{j(j+1)}\left(\frac{Z}{2}\right)^2\right)+\frac{1}{2}\ln{j(j+1)}-\psi(j+1) \nonumber \\
 &\approx& \frac{1}{2j(j+1)}\left(\frac{Z}{2}\right)^2-\frac{1}{6(j+1)^2} +\cdots,
\ea
so the leading term in the expression for $Q_j(x)$ is, in fact, just $K_0(Z)$ as found above, \eq{n=0_term}.



\subsection{The Legendre functions ${\mathsf P}_j(x)$ and ${\mathsf Q}_j(x)$ on the cut \label{Legendre_cut}}

We can continue the relation in \eq{Qj_K0} using the analytic properties of the functions involved and their behavior on the cut $x\leq 1$ to determine the Legendre functions  ${\mathsf P}_j$  and ${\mathsf Q}_j(x)$ on the cut. Thus, for $(x-1)\rightarrow e^{\pm i\pi}(1-x)$ with $x$ real, $-1<x<1$, Eqs.\ (\ref{Qj_2F1}) and (\ref{QjPjform}) give the relations (HTF1, 3.4(2), 3.4(8)) 
\ba
\label{P_on_cut}
{\mathsf P}_j^\mu(x) &=& (1/i\pi)e^{-i\pi\mu}\left[e^{-i\pi\mu/2}Q_j^\mu(x+i0) - e^{i\pi\mu/2}Q_j^\mu(x-i0)\right] \\
\label{P_on_cut2}
&=& e^{\pm i\pi\mu/2}P_j^\mu(x\pm i0) ,\\
 \label{Q_on_cut}
{\mathsf Q}_j^\mu(x) &=& (1/2)e^{-i\pi\mu}\left[e^{-i\pi\mu/2}Q_j^\mu(x+i0)+e^{i\pi\mu/2}Q_j^\mu(x-i0)\right].
\ea

Similar relations hold for the continuation of $K_\nu(Z)$ from the region $x-1>0$ to $x<1$ with $(x-1)\rightarrow (1-x)e^{\pm i\pi}$. In particular, with $Z\rightarrow e^{\pm i\pi/2}z$, $z=\sqrt{2j(j+1)(1-x)}$,
\be
\label{K(iz)}
 K_\nu\left(e^{\pm i\pi/2}z\right) = \mp \frac{i\pi}{2}e^{\mp i\pi\nu/2}\left[ J_\nu(z) \mp  i Y_\nu(z)\right]
\ee
so that on the cut
\ba
\label{K_J_relation}
J_\nu(z) &=& (i/\pi)\left[K_\nu( z+i0) -  K_\nu(z-i0) \right], \\
\label{K_Y_relation}
Y_\nu(z) &=& -(1/\pi)\left[ K_\nu( z+i0) + K_\nu(z-i0) \right].
\ea

We can use these relations and the series obtained above for $Q_j^\mu(Z)$  to construct the asymptotic Bessel function series for the Legendre functions on the cut. Thus, for $\mu=0$, the terms in \eq{Qj_K0} proportional to the Besssel functions $I_n(Z)$ have no discontinuity across the cut and drop out in the series for ${\mathsf P}_j(x)$. The discontinuities of the functions  $K_n(Z)$  give the functions $J_n(z)$, and the result reproduces the series in \eq{Pj_series} as expected, with  ${\mathsf P}_j(x)=P_j(x\pm i0)$,

In addition, using Eqs.\ (\ref{Q_on_cut}) and (\ref{K_Y_relation}), we obtain the apparently new series
\ba
{\mathsf Q}_j(x) &\approx&  -\frac{\pi}{2}Y_0(z)+J_0(Z)\left(\ln{\frac{z'}{2}}-\psi(j+1)\right) \nonumber \\
&&  +\frac{\pi}{2}\frac{1}{j(j+1)}\left[\left(\frac{z}{2}\right)^2Y_2(z)-\frac{1}{3}\left(\frac{z}{2}\right)^3Y_3(z)\right] +O\left(\frac{1}{j^2(j+1)^2}\right)
\label{Qjseries_on_cut}
\ea
for the functions ${\mathsf Q}_j(x)$ on the cut. The leading term, ${\mathsf Q}_j(x)\approx -(\pi/2)Y_0(z)$ is already known for $j\gg 1$ (HTF2 7.8(4)). 

The general relations for $\mu\not=0$ are clearly more complicated as seen above. However, the leading terms are simple, with
\ba
\label{Pmu_Qmu_leading}
{\mathsf P}_j^\mu(x) \approx \left(\frac{Z'}{2}\right)^\mu J_\mu(z),\quad {\mathsf Q}_j^\mu(x) \approx -\frac{\pi}{2}\left(\frac{Z'}{2}\right)^\mu Y_\mu(z)
\ea
%



\section{Jacobi functions and rotation coefficients \label{sec:Jacobi}}


\subsection{The functions $P^{(\alpha,\beta)}_j(x)$ \label{subsec:Jacobi_P}}

The Jacobi functions $P^{(\alpha,\beta)}_j(x)$ are defined in terms of hypergeometric functions as  \cite{szego}, Chap.\ IV,
\be
\label{Jacobi_defined}
P_j^{(\alpha,\beta)}(x)=  \frac{\Gamma(j+\alpha+1)}{\Gamma(j+1)\Gamma(\alpha+1)}\  _2F_1\left(-j,j+\alpha+\beta+1;\alpha+1; \frac{1-x}{2}\right).
\ee
For notational simplicity, we will introduce a parameter $b=\alpha+\beta+1$. 

The  hypergeometric function in \eq{Jacobi_defined}  satisfies the Barnes-type integral representation
\be
\label{Jacobi_Barnes}
 _2F_1\left(-j,j+b;\alpha+1; \frac{1-z}{2}\right)=\frac{1}{2\pi i}\int_{-i\infty}^{i\infty} ds \frac{\Gamma(-j+s)\Gamma(j+\alpha+\beta +s+1)}{\Gamma(-j)\Gamma(j+\alpha+\beta+1)}\frac{\Gamma(\alpha+1)\Gamma(-s)}{\Gamma(s+\alpha+1)}\left(-\frac{1-x}{2}\right)^s
\ee
for non-integer values of $j$, with the integration contour running to the right of the poles of  $\Gamma(-j+s)$ and to the left of the poles of $\Gamma(-s)$. Following the procedure we used in Sec. \ref{subsec:Pj} to obtain an asymptotic Bessel-function expansion for $P_j^{-\mu}(x)$, we expand the ratios of $j$-dependent gamma functions in \eq{Jacobi_Barnes} in inverse powers of $j(j+b)$ for $j$ large and find that
\ba
\frac{\Gamma(-j+s)\Gamma(j+b+s)}{\Gamma(-j)\Gamma(j+b)} &=& \left(-j(j+b)\right)^s\left[1 - \frac{1}{j(j+b)}\left(\frac{b+1}{2}s(s-1)+\frac{1}{3}s(s-1)(s-2)\right) \right.\nonumber \\
&& +\frac{1}{j^2(j+b)^2}\left(\frac{(b+1)(b+2)}{3}s(s-1)(s-2) + \frac{11+8b+b^2}{8}s(s-1)(s-2)(s-3)\right.  \nonumber \\
&& \left. +\frac{17+5 b}{30}s(s-1)(s-2)(s-3)(s-4)+\frac{1}{18}s(s-1)(s-2)(s-3)(s-4)(s-5)\right)  \nonumber \\
\label{1/j^2_expansion}
&&\left.  +\, O\left(\frac{1}{j^3(j+b)^3}\right)\right].
\ea
This series reduces to that in \eq{gamma_expansion} for $b\rightarrow 1$. 

We will use the expansion in \eq{Jacobi_Barnes}, defining a new variable $z'$ with $z'=\sqrt{2j(j+b)(1-x)}$,  and combining the factors $s(s-1)\cdots(s-k)$ with $\Gamma(-s)$ to get $(-1)^{k+1}\Gamma(-s+k+1)$. We can then shift the contour of the $s$ integration to the right to run from $k+1-i\infty$ to $k+1+i\infty$, passing just to the left of the pole at $s=k+1$. Replacing $s$ by $s'=s-k-1$ and using the result in \eq{J_(k+1)}, we find that 
\ba
P_j^{(\alpha,\beta)}(x) & \approx & \frac{ \Gamma(j+\alpha+1)}{\Gamma(j+1)}\left(\frac{z}{2}\right)^{-\alpha}\left\{ J_\alpha(z)-\frac{1}{j(j+b)}\left[\frac{b+1}{2}\left(\frac{z}{2}\right)^2 J_{\alpha+2}(z)-\frac{1}{3}\left(\frac{z}{2}\right)^3 J_{\alpha+3}(z) \right]  \right. \nonumber\\
&& -\frac{1}{j^2(j+b)^2}\left[\frac{(b+1)(b+2)}{3}\left(\frac{z}{2}\right)^3J_{\alpha+3}(z)-\frac{11+8b+b^2}{8}\left(\frac{z}{2}\right)^4J_{\alpha+4}(z) \right. \nonumber \\
\label{PJacobi_series}
&& \left. \left. +\frac{17+5b}{30}\left(\frac{z}{2}\right)^5J_{\alpha+5}(z)-\frac{1}{18}\left(\frac{z}{2}\right)^6J_{\alpha+6}(z)\right]+\cdots \right\}, \ea
with $ z=\sqrt{2j(j+b)(1-x)}$ and $b=\alpha+\beta+1$. 

We emphasize that the use of the product $j(j+b)$ in $z$ rather than the simpler $j(j+1)$ encountered in earlier sections is natural. The choice of the latter requires secondary expansions which introduce unmatched powers of $j$ and $(j+1)$ in the denominators, derivatives of the Bessel functions, and most importantly, a leading correction to the first term of order $1/j$ rather than $1/j^2$ for $j\gg 1$.

This result in \eq{PJacobi_series} reproduces the series for $P_j^{-\mu}(x)$ in \eq{Pjseries1} when $\alpha=-\beta= \mu$, $b=1$, since
\be
\label{LegendreJacobi}
P_j^{-\mu}(x) =\frac{\Gamma(j+1)}{\Gamma(j+\mu+1)}\left(\frac{x-1}{x+1}\right)^{\mu/2} P_j^{(\mu,-\mu)}(x).  
\ee

We can immediately continue the the series in \eq{PJacobi_series} from the region $-1\leq x\leq1$ to $x>1$ using the relation
\be
\label{IJacobi_series}
\left(\frac{z'}{2}\right)^{-\alpha+n}J_{\alpha+n}(z') \rightarrow (-1)^n\left(\frac{Z}{2}\right)^{-\alpha+n} I_{\alpha+n}(Z), \quad Z=\sqrt{2j(j+\alpha+\beta+1)(x-1)}
\ee
to obtain an expansion of $P_j^{(\alpha,\beta)}(x)$ for $x>1$ in terms of the hyperbolic Bessel functions $I_{\alpha+n}(Z)$ which holds for $|(x-1)/2|<1$. The limit of large $x$, with $|(x-1)/2|>1$ is of less interest, but can be treated similarly to the the large-$x$ limit of the functions which we discuss below.


\subsection{The functions $Q^{(\alpha,\beta)}_j(x)$ \label{subsec:Jacobi_Q}}

The Jacobi functions of the second kind, $Q_j^{(\alpha,\beta)}(x)$, are given by the independent hypergeometric expression \cite{szego}, 4.61.5,
\ba
\label{Qab_1}
Q_j^{(\alpha,\beta)}(x) &=& \frac{2^{j+\alpha+\beta}}{(x-1)^{j+\alpha+1}(x+1)^\beta}\frac{\Gamma(j+\alpha+1)\Gamma(j+\beta+1)}{\Gamma(2j+\alpha+\beta+2)} \,   _2F_1\left(j+1,j+\alpha+1;2j+\alpha+\beta+2;\frac{2}{1-x}\right) \\
&=&  +\frac{1}{2}\Gamma(-\alpha)\frac{\Gamma(j+\alpha+1)}{\Gamma(j+1)}\, _2F_1\left(-j,j+\alpha+\beta+1;1+\alpha;\frac{1-x}{2}\right) \nonumber \\
\label{Qab_2}
&& +\frac{1}{2}\Gamma(\alpha)\frac{\Gamma(j+\beta+1)}{\Gamma(j+\alpha+\beta+1)}\left(\frac{x-1}2\right)^{-\alpha}\left(\frac{x+1}{2}\right)^{-\beta}\, _2F_1\left(-j-\alpha-\beta,j+1;1-\alpha:\frac{1-x}{2}\right) \\
\label{Qab_3}
&=& \frac{\pi}{2\sin{\pi\alpha}}\left[-P_j^{(\alpha,\beta)}(x)+\left(\frac{x-1}{2}\right)^{-\alpha}\left(\frac{x+1}{2}\right)^{-\beta}P_{j+\alpha+\beta}^{(-\alpha,-\beta)}(x)\right].
\ea
This function satisfies the same recurrence relations as $P^{(\alpha,\beta)}(x)$.

We will assume initially that $x>1$ with $(x-1)/2<1$. The Jacobi function $P_j^{(\alpha,\beta)}(x)$ in \eq{Qab_2} can then be approximated in terms of the Bessel functions $I_{\alpha+n}(Z)$ using Eqs.\ (\ref{PJacobi_series}) and (\ref{IJacobi_series}).

We obtain the expansion for the second Jacobi function in \eq{Qab_3} by making the substitutions $j\rightarrow j+\alpha+\beta,\ b\rightarrow 1$, in the expansion in \eq{PJacobi_series}. This results in changes in the $j$ dependence of both the coefficients and  the arguments of the Bessel functions in Eqs.\ (\ref{PJacobi_series}) and (\ref{IJacobi_series}).  The combination of the two types of terms with the coefficients in \eq{PJacobi_series} gives the general asymptotic expansion for $Q_j^{(\alpha,\beta)}(x)$. It necessarily involves combinations of Bessel functions with different arguments unless $\alpha+\beta=0$, the case encountered with the Lagendre functions.
 
Since $(j+\alpha+\beta)(j+1)=j(j+\alpha+\beta+1)+(\alpha+\beta)$, we can use secondary expansions in powers of $(\alpha+\beta)/j(j+\alpha+\beta+1)$ to recover the the original argument used in the case of $P_j^{(\alpha,\beta)}(x)$, but are necessarily left with new terms that do not appear in the original series. For example, the argument $Z$ of the Bessel functions  in the asymptotic series for $P_j^{(\alpha,\beta)}(x)$ in \eq{IJacobi_series} is changed in the series for $P_{j+\alpha+\beta}^{(-\alpha,-\beta)}(x)$ to
\ba
 \sqrt{2(j +\alpha+\beta)(j+1)(x-1)} &=& Z\sqrt{1+\left(\alpha+\beta)/j(j+\alpha+\beta+1)\right)} \nonumber \\
\label{delta_z}
&=& Z+\frac{Z}{2}\frac{\alpha+\beta}{j(j+\alpha+\beta+1)}+\cdots, 
 \ea
 and we find that
 \ba
 \left(\frac{Z}{2}\right)^{\alpha+n}I_{-\alpha+n}(Z) &\rightarrow&  \left(\frac{Z}{2}\right)^{\alpha+n}I_{-\alpha+n}(Z)  \nonumber \\
 \label{I(Z+DeltaZ}
 &+&  \frac{\alpha+\beta}{j(j+\alpha+\beta+1)}\left[n\left(\frac{Z}{2}\right)^{\alpha+n}I_{-\alpha+n}(Z)+\left(\frac{Z}{2}\right)^{\alpha+n+1}I_{-\alpha+n+1}(Z)\right]+\cdots
 \ea
 in the analog of \eq{PJacobi_series}.
 
 Using this expansion and following our earlier procedure for the Legendre functions, we will approximate the Jacobi functions in \eq{Qab_3} by  hyperbolic Bessel functions using the expansion in \eq{PJacobi_series} and the relation in \eq{IJacobi_series} for $x>1$, and then express the function $I_{-\alpha}(Z)$ which appears in terms of $K_{\alpha}(Z)$. The leading terms in the resulting expression for $Q_j^{\alpha,\beta}(x)$ are
 \ba
 Q_j^{(\alpha,\beta)}(x) &\approx& \frac{\pi}{2\sin{\pi\alpha}}\frac{\Gamma(j+\alpha+1)}{\Gamma(j+1)}\left[\frac{\Gamma(j-\alpha+1)}{\Gamma(j+\alpha+1)}\left(\frac{x-1}{2}\right)^{-\alpha}\left(\frac{x+1}{2}\right)^{-\beta}\left(\frac{Z}{2}\right)^\alpha I_{-\alpha}(Z) - \left(\frac{Z}{2}\right)^{-\alpha} I_\alpha(Z)\right] \nonumber \\
 &\approx& \frac{\pi}{2\sin{\pi\alpha}}\frac{\Gamma(j+\alpha+1)}{\Gamma(j+1)}\left(\frac{Z}{2}\right)^{-\alpha}\left[I_{-\alpha}(Z)\left(1+\frac{\alpha(\alpha+\beta)}{j+1}+\cdots\right) - I_\alpha(Z)+O\left(\frac{1}{(j+1)^2}\right)\right] \nonumber \\
 &=& \frac{\Gamma(j+\alpha+1)}{\Gamma(j+1)}\left[\left(\frac{Z}{2}\right)^{-\alpha}K_{\alpha}(Z)\left(1+\frac{\alpha(\alpha+\beta)}{j+1}+\cdots\right)\right. \nonumber \\
 \label{Qjab_series}
 &&\quad\left.+ \frac{\pi}{2\sin{\pi\alpha}}\frac{\alpha(\alpha+\beta)}{j+1}\left(\frac{Z}{2}\right)^{-\alpha} I_\alpha(Z)+\cdots\right].
 \ea
The terms of order $1/(j+1)$ in \eq{Qjab_series} arise from the expansion of the ratio of the  prefactor to $I_{-\alpha}(Z)$ and the factor $\left((j(j+\alpha+\beta+1)\right)^{-\alpha}$ needed  to convert $(\left(x-1)/2\right)^{-\alpha}$ to $(Z/2)^{-2\alpha}$. This expansion is not necessary, but leaves a complicated overall coefficient for $K_\alpha(Z)$ if not done.  We choose to show its leading dependence on $j$ explicitly.

We do not give terms in the expansion $Q_j^{(\alpha,\beta)}(x)$ of order $1/(j+1)^2$ or higher, though they can be obtained from the general expansion using results given previously.

Our results above hold for $|x-1|/2<1$. However, the functions $Q_j^{(\alpha,\beta)}(x)$ are also needed in various contexts for $|x-1|/2>1$. We can again use the methods developed above to obtain a Bessel function approximation in this region. 
 We start with the expression in \eq{Qab_1} and use a Barnes representation of the hypergeometric function to rewrite this as
\ba
Q_j^{(\alpha,\beta)}(x) &=& \frac{1}{2}\left(\frac{x+1}{2}\right)^{-j-\alpha-1}\left(\frac{x+1}{2}\right)^{-\beta}\Gamma(j+\alpha+1)\Gamma(j+\beta+1) \nonumber \\
\label{Qj_large1}
&& \times\frac{1}{2\pi i}\int_{-i\infty}^{i\infty}ds\frac{\Gamma(j+s+1)\Gamma(j+\alpha+s+1)}{\Gamma(j+1)\Gamma(j+\alpha+1)}\frac{\Gamma(-s)}{\Gamma(2j+\alpha+\beta+s+2)}\left(\frac{2}{x-1}\right)^s.
\ea
 Expanding the first ratio of gamma functions inside the integral assuming that $j\gg s$ on the relevant parts of the contour and using Stirling's approximation, we obtain
 \ba
 \frac{\Gamma(j+s+1)\Gamma(j+\alpha+s+1)}{\Gamma(j+1)\Gamma(j+\alpha+1)} &\approx& \left(j_1j_2\right))^s\left[1+\frac{1}{2}s(s-1)\left(\frac{1}{j_1}+\frac{1}{\j_2}+\frac{1}{j_1j_2}\right) + \frac{1}{3}s(s-1)(s-2)\left(\frac{1}{j_1^2}+\frac{1}{j_2^2}+\frac{3}{j_1j_2}\right)  \right. \nonumber \\
 \label{Qj_large2}
 && \left. + \frac{1}{8}s(s-1)(s-2)(s-3)\left(\frac{1}{j_1}+\frac{1}{j_2}\right)^2 + O\left(\frac{1}{j^3}\right)\right],
 \ea
 where $j_1=j+1,\ j_2=j+\alpha+1$. This expansion agrees with \eq{gamma_expansion} for $j_1=-j,\ j_2=j+1$.
 
 Using this expansion in \eq{Qj_large1} and then using the representation for the Bessel functions in \eq{Bessel_int}, we find that
 \ba
Q_j^{(\alpha,\beta)}(x) &\approx&  \frac{1}{2}\left(\frac{x-1}{2}\right)^{-j-\alpha-1}\left(\frac{x+1}{2}\right)^{-\beta}\Gamma(j+\alpha+1)\Gamma(j+\beta+1) \nonumber \\
&& \times\left(\frac{Z''}{2}\right)^{-\nu}\left\{J_\nu(Z'')+\frac{1}{2}\left(\frac{1}{j_1}+\frac{1}{\j_2}+\frac{1}{j_1j_2}\right)\left(\frac{Z''}{2}\right)^2 J_{\nu+2}(Z'') \right. \nonumber \\
\label{Qj_large3}
&&\left.  -\frac{1}{3}\left(\frac{1}{j_1^2}+\frac{1}{j_2^2}+\frac{3}{j_1j_2}\right)\left(\frac{Z''}{2}\right)^3J_{\nu+3}(Z'') +\frac{1}{8}\left(\frac{1}{j_1}+\frac{1}{j_2}\right)^2\left(\frac{Z''}{2}\right)^4J_{\nu+4}(Z'')+\cdots\right\},
\ea
where $Z''=\sqrt{8j_1j_2/(x-1)}=\sqrt{8(j+1)(j+\alpha+1)/(x-1)}$ and $\nu=2j+\alpha+\beta+1$. The series in the brackets is asymptotic, but potentially useful for $j$ large and $Z''$ fixed and $O(1)$ or smaller. The overall result reproduces the expected behavior limiting behavior of $Q_j^{(\alpha,\beta)}(x)$ for $x\rightarrow\infty$.

We can obtain an an alternative form of this expression which is somewhat better for $x$ not too large by using a standard transformation formula for the hypergeometric function in \eq{Qab_1} to convert the argument from $2/(x-1)$ to $2/(x+1)$, and then following the same procedure. The result is
 \ba
 \label{Qab_alt}
Q_j^{(\alpha,\beta)}(x) &=& \frac{1}{2}\left(\frac{x+1}{2}\right)^{-j-\alpha-\beta-1}\frac{\Gamma(j+\alpha+1)\Gamma(j+\beta+1)}{\Gamma(2j+\alpha+\beta+2)}\, _2F_1\left(j+1,j+\beta+1;2j+\alpha+\beta+2;\frac{2}{x+1}\right)  \\
&\approx&  \frac{1}{2}\left(\frac{x+1}{2}\right)^{-j-\alpha-\beta-1}\Gamma(j+\alpha+1)\Gamma(j+\beta+1) \nonumber \\
&& \times\left(\frac{Y}{2}\right)^{-\nu}\left\{I_\nu(Y)+\frac{1}{2}\left(\frac{1}{j_1}+\frac{1}{\j_2}+\frac{1}{j_1j_2}\right)\left(\frac{Y}{2}\right)^2 I_{\nu+2}(Y) \right. \nonumber \\
\label{Qj_large4}
&&\left.  +\frac{1}{3}\left(\frac{1}{j_1^2}+\frac{1}{j_2^2}+\frac{3}{j_1j_2}\right)\left(\frac{Y}{2}\right)^3I_{\nu+3}(Y) +\frac{1}{8}\left(\frac{1}{j_1}+\frac{1}{j_2}\right)^2\left(\frac{Y}{2}\right)^4I_{\nu+4}(Y)+\cdots\right\},
\ea
where $Y=\sqrt{8j_1j_2/(x+1)}=\sqrt{8(j+1)(j+\beta+1)/(x+1)}$ and $\nu=2j+\alpha+\beta+1$.

The hypergeometric expressions for $Q_j^{(\alpha,\beta)}(x)$ in Eqs.\ (\ref{Qab_1}) and (\ref{Qab_2})  hold, respectively, outside and inside a circle of radius 2 centered at $x=1$. The results in 
\eq{Qjab_series} are, similarly, useful inside the circle with $|x-1|/2$ small, while Eqs.\ (\ref{Qj_large1}) and (\ref{Qj_large3}) hold outside, for $|x-1|/2$ large. Neither approximation is useful near the circle, in particular near the boundary point at $x=3$ for $x$ real. The representation in \eq{Qab_alt} is valid in the boundary region, but the asymptotic expansion still requires  $|x|\gg 1$. 


\subsection{The Jacobi functions ${\mathsf P}_j^{(\alpha,\beta)}(x)$ and ${\mathsf Q}_j^{(\alpha,\beta)}(x)$ on the cut \label{subsecJacobi_on_cut}}

It is simple to obtain the Bessel function approximations for the Jacobi functions on the cut $-1< x < 1$ using the defining relations \cite{szego} (4.62.8,\, 9)
\ba
\label{Pcut}
{\mathsf P}_j^{(\alpha,\beta)}(x) &=& \frac{i}{\pi}\left(e^{i\pi\alpha}Q_j^{(\alpha,\beta)}(x+i0) - e^{-i\pi\alpha}Q_j^{(\alpha,\beta)}(x-i0)\right) \\
&=& P_j^{(\alpha,\beta)}(x\pm i0), \\
\label{Qcut}
{\mathsf Q}_j^{(\alpha,\beta)}(x) &=& \frac{1}{2}\left(e^{i\pi\alpha}Q_j^{(\alpha,\beta)}(x+i0) + e^{-i\pi\alpha}Q_j^{(\alpha,\beta)}(x-i0)\right). 
\ea

The result for ${\mathsf P}_j^{(\alpha,\beta)}(x)$ follows from \eq{PJacobi_series},
\ba
{\mathsf P}_j^{(\alpha,\beta)}(x) & \approx & \frac{ \Gamma(j+\alpha+1)}{\Gamma(j+1)}\left(\frac{z}{2}\right)^{-\alpha}\left\{ J_\alpha(z)-\frac{1}{j(j+b)}\left[\frac{b+1}{2}\left(\frac{z}{2}\right)^2 J_{\alpha+2}(z)-\frac{1}{3}\left(\frac{z}{2}\right)^3 J_{\alpha+3}(z) \right] +\cdots \right\},
\label{Pcut2}
\ea
$b=\alpha+\beta+1$. We obtain the leading terms in the expansion for ${\rm Q}_J^{\alpha,\beta)}(x)$ using \eq{Qjab_series} and the relations in \eq{K(iz)},
\ba
{\mathsf Q}_j^{(\alpha,\beta)}(x) &\approx&   -\frac{\pi}{2}\frac{\Gamma(j+\alpha+1)}{\Gamma(j+1)}\left\{\left(\frac{z}{2}\right)^{-\alpha}Y_{\alpha}(z)\left(1+\frac{\alpha(\alpha+\beta)}{j+1}+\cdots\right)\right. \nonumber \\
 \label{Qjab_seriescut}
&& \quad\left.-\frac{\cos{\pi\alpha}}{\sin{\pi\alpha}}\frac{\alpha(\alpha+\beta)}{j+1}\left(\frac{z}{2}\right)^{-\alpha} J_\alpha(z)+\cdots\right\}.
 \ea
We note that the correction terms of order $1/(j+1)$ vanish for $\beta=-\alpha$, the case encountered for the Legendre functions $Q_j^\mu(x)$ where the corrections to the leading term are order $1/j(j+1)$.


\subsection{The rotation coefficients  $d_{m'm}^j(x)$ and $e_{m'm}^j(x)$ \label{rotation_coeff}}

An important application of our results involves the rotation coefficients $\mathsf d_{m'm}^j(x)=\langle j,m'| e^{i\theta J_y}|j,m\rangle$, $x=\cos{\theta}$, which we define here using Edmonds' conventions for passive coordinate rotations \cite{Edmonds}. We will also consider the associated functions of the second kind ${\mathsf e}_{m'm}^j(x)$ \cite{LD_chiu1965}.  In such physical applications,  $j$, $m'$, and $m$ are all either integers or half-integers, and can be arranged using the symmetries of the functions so that $j\geq m'\geq |m|$.  Then, following Edmonds,
\be
\label{rotation_ds}
{\mathsf d}_{m',m}^j(\cos{\theta}) = \left[\frac{(j+m')!(j-m')!}{(j+m)!(j-m)!}\right]^{1/2}\left(\sin{\frac{\theta}{2}}\right)^{(m'-m)}\left(\cos{\frac{\theta}{2}}\right)^{(m'+m)}P_{j-m'}^{(m'-m),m'+m)}(\cos{\theta}).
\ee
For active rotations of the system, $\theta\rightarrow-\theta$, and the rotation functions defined in that convention \cite{Rose} acquire a factor $(-1)^{m'-m}$ on the right-hand side when written as above, with $\sin{(\theta/2)}$ taken as the positive square root of $(1-\cos{\theta})/2$. 

The functions ${\mathsf d}^j_{m',m}(x)$ are needed to describe the rotations of the eigenstates of angular momentum \cite{Edmonds}, and  the scattering of particles with spin \cite{jacob_wick}. In particular, the ${\mathsf d}^j_{\lambda,\mu}(x)$ appear as the angular coefficients in the partial-wave expansions of scattering amplitudes $A_{\lambda_c,\lambda_d;\lambda_a,\lambda_b}$ for processes $a+b\rightarrow c+d$, with  $\theta$ the polar scattering angle, and $m'$ and $m$ the total initial and final particle helicities $m'=\lambda_a-\lambda_b$ and $m=\lambda_c-\lambda_d$. These indices can be rearranged and relabelled as necessary using symmetries so as to satisfy the inequalities  above \cite{jacob_wick}. The rotation functions of the second kind, $e^j_{m',m}(x)$, similarly appear as coefficient functions when the partial-wave scattering amplitudes are expressed as integrals over the weight functions in dispersion relation representations of the scattering amplitude in $x=\cos{\theta}$.

We will begin here with the general definitions
\ba
\label{dj}
d_{m',m}^j(x) &=& \left[\frac{\Gamma(j+m'+1)\Gamma(j-m'+1)}{\Gamma(j+m+1)\Gamma(j-m+1)}\right]^{1/2}\left(\frac{x-1}{2}\right)^{\frac{1}{2}(m'-m)}\left(\frac{x+1}{2}\right)^{\frac{1}{2}(m'+m)}P_{j-m'}^{(m'-m),m'+m)}(x) \\
&=& \left[\frac{\Gamma(j+m'+1)\Gamma(j-m+1}{\Gamma(j-m'+1)\Gamma(j+m+1)}\right]^{1/2}\left(\frac{x-1}{2}\right)^{\frac{1}{2}(m'-m)}\left(\frac{x+1}{2}\right)^{\frac{1}{2}(m'+m)} \nonumber \\
\label{dj2}
&& \times\frac{1}{\Gamma(m'-m+1)}\, _2F_1\left(-j+m',j+m'+1;m'-m+1;\frac{1-x}{2}\right) \\
\label{ej}
e_{m',m}^j(x) &=& \left[\frac{\Gamma(j+m'+1)\Gamma(j-m'+1)}{\Gamma(j+m+1)\Gamma(j-m+1)}\right]^{1/2}\left(\frac{x-1}{2}\right)^{\frac{1}{2}(m'-m)}\left(\frac{x+1}{2}\right)^{\frac{1}{2}(m'+m)}Q_{j-m'}^{(m'-m),m'+m)}(x) \\
&=& \frac{1}{2}\left[\Gamma(j+m'+1)\Gamma(j+m+1)\Gamma(j-m'+1)\Gamma(j-m+1)\right]^{1/2}\left(\frac{x-1}{2}\right)^{-j+\frac{1}{2}(m'+m)-1}\left(\frac{x+1}{2}\right)^{-\frac{1}{2}(m'+m)} \nonumber \\
\label{ej2}
&& \times \frac{1}{\Gamma(2j+2)}\, _2F_1\left(j-m'+1,j-m+1;2j+2;\frac{2}{1-x}\right),
\ea
with no restrictions on $j,\,m',$ and $m$, and $x$ a complex variable. 

There are branch points in $d_{m',m}^j(x)$ and $e_{m',m}^j(x)$ at $x=1$ and $x=-1$. The functions are analytic in the complex $x$ plane cut from $x=1$ and $x=-1$ to $-\infty$. This is in accord with standard treatments of the Jacobi functions \cite{szego}, and with our earlier treatment of the rotation functions in \cite{LD_chiu1965}, but differs from the choice of cuts used by Andrews and Gunson in \cite{AndrewsGunson} in their discussion of the analytic properties of these functions.

The rotation coefficients as defined in \eq{rotation_ds}  involve the Jacobi functions ``on the cut'' with $x=\cos{\theta}$ real and $-1\leq x\leq1$. We will extract these functions following our treatment of the Jacobi functions ${\mathsf P}_j^{(\alpha,\beta}(x)$ and ${\mathsf Q}_j^{(\alpha,\beta)}(x)$ on the cut.  From \eq{Qab_3}, 
\ba
e^j_{m',m}(x) &=& \left[\frac{\Gamma(j+m'+1)\Gamma(j-m'+1)}{\Gamma(j+m+1)\Gamma(j-m+1)}\right]^{1/2} \frac{\pi}{2\sin{\pi(m'-m)}} \nonumber \\
&& \left\{-\left(\frac{x-1}{2}\right)^{\frac{1}{2}(m'-m)}\left(\frac{x+1}{2}\right)^{\frac{1}{2}(m'+m)} P_{j-m'}^{(m'-m,m'+m)}(x) \right. \nonumber \\
\label{ej_alt}
&& + \left. \left(\frac{x-1}{2}\right)^{-\frac{1}{2}(m'-m)}\left(\frac{x+1}{2}\right)^{-\frac{1}{2}(m'+m)} P_{j+m'}^{(-m'+m,-m'-m)}(x)\right\}.
\ea
Combining the functions $e^j_{m',m}(x\pm i0)$ for $x$ real, $-1<x<1$, as in \eq{Pcut} to extract the Jacobi function ${\mathsf P}_{j-m'}^{(m'-m,m'+m)}(x)$ on the cut, we find that an appropriate definition of the functions ${\mathsf d}^j_{m',m}(x)$ on the cut is 
\ba
\label{dj_defined1}
{\mathsf d}^j_{m'm}(x) &=& \frac{i}{\pi}\left(e^{i\frac{\pi}{2}(m'-m)}e^j_{m',m}(x+i0) - e^{-i\frac{\pi}{2}(m'-m)}e^j_{m',m}(x-i0)\right) \\
\label{dj_defined2}
&=& \left[\frac{\Gamma(j+m'+1)\Gamma(j-m'+1)}{\Gamma(j+m+1)\Gamma(j-m+1)}\right]^{1/2}\left(\frac{1-x}{2}\right)^{\frac{1}{2}(m'-m)}\left(\frac{1+x}{2}\right)^{\frac{1}{2}(m'+m)}{\mathsf P}_{j-m'}^{(m'-m),m'+m)}(x) .
\ea
This agrees with the definition of the rotation coefficients in \eq{rotation_ds} for $x=\cos{\theta}$.

Similarly, we find from Eqs.\ (\ref{ej}) and (\ref{Qcut}) that the rotation coefficients of the second kind ${\mathsf e}^j_{m',m}(x)$ are given on the cut in terms of the functions ${\mathsf Q}^{(m'-m,m'+m)}_{j-m'}(x)$ by
\ba
\label{ej_defined1}
{\mathsf e}^j_{m',m} &=& \frac{1}{2}\left(e^{i\frac{\pi}{2}(m'-m)}e^j_{m',m}(x+i0) + e^{-i\frac{\pi}{2}(m'-m)}e^j_{m',m}(x-i0)\right) \\
\label{ej_defined2}
&=&  \left[\frac{\Gamma(j+m'+1)\Gamma(j-m'+1)}{\Gamma(j+m+1)\Gamma(j-m+1)}\right]^{1/2}\left(\frac{1-x}{2}\right)^{\frac{1}{2}(m'-m)}\left(\frac{1+x}{2}\right)^{\frac{1}{2}(m'+m)}{\mathsf Q}_{j-m'}^{(m'-m),m'+m)}(x).
\ea

The symmetries of these functions for physical, integer or half integer, values of $m'$ and $m$ follow from those of the general functions $e^j_{m',m}$. We can easily see from the properties of the hypergeometric function in \eq{ej2} that $e^j_{m,m'}(x)=e^j_{-m',-m}(x)=e^j_{m',m}(x)$ for general $m',\,m$. The factors $e^{\pm i\frac{\pi}{2}(m'-m)}$ in Eqs.\ (\ref{dj_defined1}) and (\ref{ej_defined1}) change the symmetries of the two terms in these relations by factors $e^{\pm i\pi(m'-m)} = (-1)^{(m'-m)}$ when $m'$ and $m$ are exchanged or changed in sign,
so that ${\mathsf d}^j_{m,m'}(x)={\mathsf d}^j_{-m',-m}(x)=(-1)^{(m'-m)}{\mathsf d}^j_{m',m}(x)$ and ${\mathsf e}^j_{m,m'}(x)={\mathsf e}^j_{-m',-m}(x)=(-1)^{(m'-m)}{\mathsf e}^j_{m',m}(x)$.

We can also, for physical values of $m'$ and $m$, insert overall factors $(-1)^{(m'-m)}$ in Eqs.\ (\ref{dj_defined2}) and (\ref{ej_defined2}) to convert from the Edmonds convention for coordinate rotations  \cite{Edmonds} to the Rose convention for active rotations of the system \cite{Rose}.

We can now use our previous results to obtain asymptotic Bessel function approximations for the rotation functions. From \eq{Pcut2}
\ba
{\mathsf d}^j_{m'm}(x) &\approx& \frac{\Gamma(j+m'+1)\Gamma(j-m+1)}{\Gamma(j-m'+1)\Gamma(j+m+1)}\left(\frac{1-x}{2}\right)^{\frac{1}{2}(m'-m)}\left(\frac{1+x}{2}\right)^{\frac{1}{2}(m'+m)} \left(\frac{z}{2}\right)^{-(m'-m)} \nonumber \\
\label{dcutBessel}
&\times& \left\{ J_{m'-m}(z)-\frac{1}{(j-m')(j+m'+1)}\left[(m'+1)\left(\frac{z}{2}\right)^2 J_{m'-m+2}(z)-\frac{1}{3}\left(\frac{z}{2}\right)^3 J_{m'-m+3}(z) \right] +\cdots \right\},
\ea
where $z=\sqrt{2(j-m')(j+m'+1)(1-x)}=\sqrt{2[j(j+1)-m'(m'+1)](1-x)}$. \footnote{A different argument with the term $m'(m'+1)$ in the coefficient of $(1-x)$ replaced by $m'(m'+1)/2 +m(m+1)/2)$ was suggested in \cite{LD_chiu1965}. This is incorrect. It was motivated by expanding the factor $\left((1+x)/2\right)^{(m'+m)/2}$ multiplying the Jacobi function and then requiring that the expansion of the Bessel function match the result to order $(1-x)$. The proper result given here also includes a term of order $(1-x)$ from the first correction term in the series in \eq{dcutBessel}. The possibility of such a correction in a systematic expansion was not considered earlier.} 

Similarly, from \eq{Qjab_seriescut},
\ba
{\mathsf e}^j_{m',m}(x) &\approx& -\frac{\pi}{1}\frac{\Gamma(j+m'+1)\Gamma(j-m+1)}{\Gamma(j-m'+1)\Gamma(j+m+1)}
\left(\frac{1-x}{2}\right)^{\frac{1}{2}(m'-m)}\left(\frac{1+x}{2}\right)^{\frac{1}{2}(m'+m)} \left(\frac{z}{2}\right)^{-(m'-m)} \nonumber \\
\label{ecutBessel}
&\times& \left\{ Y_{m'-m}(z)\left(1+\frac{2m'(m'-m)}{j-m'+1}+\cdots\right) - \frac{\cos{\pi(m'-m)}}{\sin{\pi(m'-m)}}\cdot\frac{2m'(m'-m)}{j-m'+1} J_{m'-m}(z)+\cdots\right\}.
\ea

The more important applications of the functions of the second kind involve $e^j_{m',m}(x)$ for $x>1$. For $(x-1)/2\gg 1$, 
Eqs.\ (\ref{Qj_large3}) and (\ref{ej}) give
\ba
e^j_{m',m}(x) &\approx& \frac{1}{2}\left[\Gamma(j+m'+1)\Gamma(j+m+1)\Gamma(j-m'+1)\Gamma(j-m+1)\right]^{1/2}\left(\frac{x-1}{2}\right)^{-j+\frac{1}{2}(m'+m)-1}\left(\frac{x+1}{2}\right)^{-\frac{1}{2}(m'+m)} \nonumber \\
\label{ej_large}
&& \times \left(\frac{Z''}{2}\right)^{-\nu}\left\{J_\nu(Z'')+\frac{1}{2}\left(\frac{1}{j_1}+\frac{1}{\j_2}+\frac{1}{j_1j_2}\right)\left(\frac{Z''}{2}\right)^2 J_{\nu+2}(Z'') \right. \nonumber \\
\label{ej_large2}
&& \left.  -\frac{1}{3}\left(\frac{1}{j_1^2}+\frac{1}{j_2^2}+\frac{3}{j_1j_2}\right)\left(\frac{Z''}{2}\right)^3J_{\nu+3}(Z'') +\frac{1}{8}\left(\frac{1}{j_1}+\frac{1}{j_2}\right)^2\left(\frac{Z''}{2}\right)^4J_{\nu+4}(Z'')+\cdots\right\},
\ea
where $j_1=j-m'+1$, $j_2=j-m+1$, $Z''=\sqrt{8(j-m'+1)(j-m+1)/(x-1)}$ and $\nu=2j+1$. 

The leading term in this rather curious looking expression was encountered in \cite{LD_chiu1965} in a calculation of the partial-wave scattering amplitudes from a Fourier-Bessel representation of the dispersion relation for the  full amplitude for the scattering of particles with spin. It provided the connection from that general representation back to usual partial-wave expansion, where the partial-wave amplitudes can be expressed as integrals of $e^j_{m',m}(x)$ over the dispersive weight function.


\begin{acknowledgments}
The author would  like to thank the Aspen Center for Physics for its hospitality while parts of this work were done, and for its partial support of the work under NSF Grant No. 1066293.
\end{acknowledgments}


\end{document}